# Terahertz pulse generation with binary phase control in non-linear InAs metasurface

*Hyunseung Jung[1,2,*], Lucy L Hale[3], Sylvain D. Gennaro[1,2], Jayson Briscoe[1,2], Prasad P. Iyer[1,2], Chloe F. Doiron[1,2], C. Thomas Harris[1,2], Ting Shan Luk[1,2], Sadhvikas J Addamane[1,2], John L Reno[1,2], Igal Brener[1,2], Oleg Mitrofanov[1,3,*]*

[1]Sandia National Laboratories, Albuquerque, New Mexico 87123, USA

[2]Center for Integrated Nanotechnologies, Sandia National Laboratories, Albuquerque, New Mexico 87123, USA

[3]University College London, Electronic and Electrical Engineering, London WC1E 7JE, UK

ABSTRACT

The effect of terahertz (THz) pulse generation has revolutionized broadband coherent spectroscopy and imaging at THz frequencies. However, THz pulses typically lack spatial structure, whereas structured beams are becoming essential for advanced spectroscopy applications. Non-linear optical metasurfaces with nanoscale THz emitters can provide a solution by defining the beam structure at the generation stage. We develop a non-linear InAs metasurface consisting of nanoscale optical resonators for simultaneous generation and structuring of THz beams. We find that THz pulse generation in the resonators is governed by optical rectification. It is more efficient than in ZnTe crystals, and it allows us to control the pulse polarity and amplitude, offering a platform for realizing binary-phase THz metasurfaces. To illustrate this capability, we demonstrate an InAs metalens, which simultaneously generates and focuses THz pulses. The control of spatiotemporal structure using nanoscale emitters opens doors for THz beam engineering and advanced spectroscopy and imaging applications.

TEXT

Many material systems emit a single-cycle electromagnetic pulse of terahertz (THz) frequencies when excited by a femtosecond optical pulse due to rapidly induced material polarization.[1] Waveforms of the THz pulses have shed light on the rich underlying physics of induced polarization, spanning from shift currents,[2-3] charge separation in coupled quantum wells [4] and 2D monolayer heterostructures[5] to the inverse spin-Hall [6] and photon drag effects.[7] Broadband THz pulses have also revolutionized spectroscopy and imaging in the THz frequency range.[8] Advanced spectroscopy and imaging now also require control of the spatial structure of THz pulses [9-11] and their waveform.[12] Structured THz beams can be formed with the help of phase plates, spatial light modulators, polarizers and lenses,[13-19] however for broadband THz pulses, this approach is limited in bandwidth and efficiency. Instead, the spatiotemporal control can be realized at the THz generation stage. This method would not only overcome the bandwidth and efficiency limitations but eliminate the need for additional THz elements entirely.

One way to define the spatiotemporal beam structure is by generation of THz pulses with desired polarization, amplitude and phase at the subwavelength scale. Recently introduced non-linear optical metasurfaces are ideally suited for this task: they can convert optical excitation into new frequencies (e.g. harmonic generation and photoluminescence) and simultaneously steer the emission into a desired direction.[20-26] Nanoscale metallic split-ring resonators excited by optical pulses were shown to generate THz pulses,[22] and also to steer and focus THz waves.[23-26] Metallic resonators however showed relatively low efficiency and their use is limited to moderate excitation intensities.[24, 28]

Alternatively, all-dielectric metasurfaces could provide a platform for controlling the phase and amplitude of THz pulses potentially with higher conversion efficiency and higher damage threshold.[29-35] Furthermore, a broad range of THz generation mechanisms can be accessed with all-dielectric metasurfaces. In particular, the optical rectification effect due to shift currents in a GaAs metasurface was demonstrated recently to switch the polarity of generated THz pulses, introducing a $\pi$-phase shift.[35]

InAs is one of the most robust and efficient THz emitters known to date.[36-41] However, controlling the polarity of THz pulses generated in InAs crystals is difficult without modifying the material. Although InAs belongs to the same symmetry group as GaAs ($\bar{4}3m$), other mechanisms, specifically the photogenerated transient currents, tend to dominate in the THz generation process. Therefore, for InAs metasurfaces to enable THz generation with complex spatiotemporal structure it is essential not only to activate optical rectification, but also to control all other generation mechanisms. Here, we report on an InAs metasurface consisting of nanoscale resonators, which serve as nano-scale THz emitters with the optical rectification effect playing the dominant role. We find that the polarity of THz pulses generated in the InAs resonators is reversed in comparison to a uniform InAs layer of the same thickness, while their amplitudes are similar. This remarkable combination offers an opportunity to engineer the spatial profile of radiated THz pulses. The optical-to-THz conversion efficiency in the InAs metasurface exceeds that of a bulk ZnTe crystal. As a proof of principle, we demonstrate THz beam generation and simultaneous focusing using an InAs metasurface lens.

THz generation in bulk InAs crystals has been attributed mainly to photocurrent mechanisms,[36-39] which offer limited options for controlling polarity of emitted THz pulses. In contrast, the volume optical rectification effect depends sensitively on the material crystallographic orientation and the excitation polarization.[35,42] While the volume non-linear contribution in InAs has been overshadowed by the photocurrent mechanisms, this polarization dependence can enable control of the THz generation process.

To enhance the volume optical rectification mechanism and suppress the photocurrent mechanisms, we designed a metasurface consisting of electrically-isolated rectangular nanoscale resonators. The second-order non-linear tensor, $\chi^{(2)}$, provides us with the following guiding principle: when the incident electric field vector is aligned along any of the [111]- or [110]-group directions in an InAs crystal, we expect it to induce the strongest non-linear THz polarization. For the most widely used InAs with (001) surface orientation, when the crystal is tilted by ~45-60 degrees with respect to the optical axis and rotated by 45 degrees with respect to the [001] direction, as illustrated in Figure 1, the electric field vector of the incident *p*-polarized excitation is aligned close to the [–111] direction. For the *s*-polarized excitation and the same crystal orientation, the electric field is aligned along the [110] direction. Therefore, the configuration in Figure 1 provides a convenient crystal orientation for activating the second order non-linearity using both the *p*-polarized and *s*-polarized excitation.

In a metasurface, to apply this guiding principle we need to consider the field distribution of metasurface modes instead of the incident field. In a rectangular resonator, the lowest order modes are the electric dipole (ED) and magnetic dipole (MD) modes, and their wavelengths can be adjusted by the resonator size and the spacing between neighboring resonators. For the resonator orientation in Figure 1, where *x*, *y* and *z* represent the resonator axes, we selected a resonator size and spacing which result in the fundamental in-plane ED modes, $ED_x$ and $ED_y$, centered at 770 nm (see Supplementary Information (SI) for details). Each of these two modes induces a component of non-linear polarization $P_{THz,p}$ aligned along the vertical axis. Therefore, when these modes are excited by an incident optical beam, a *p*-polarized THz pulse will be emitted in the forward direction.

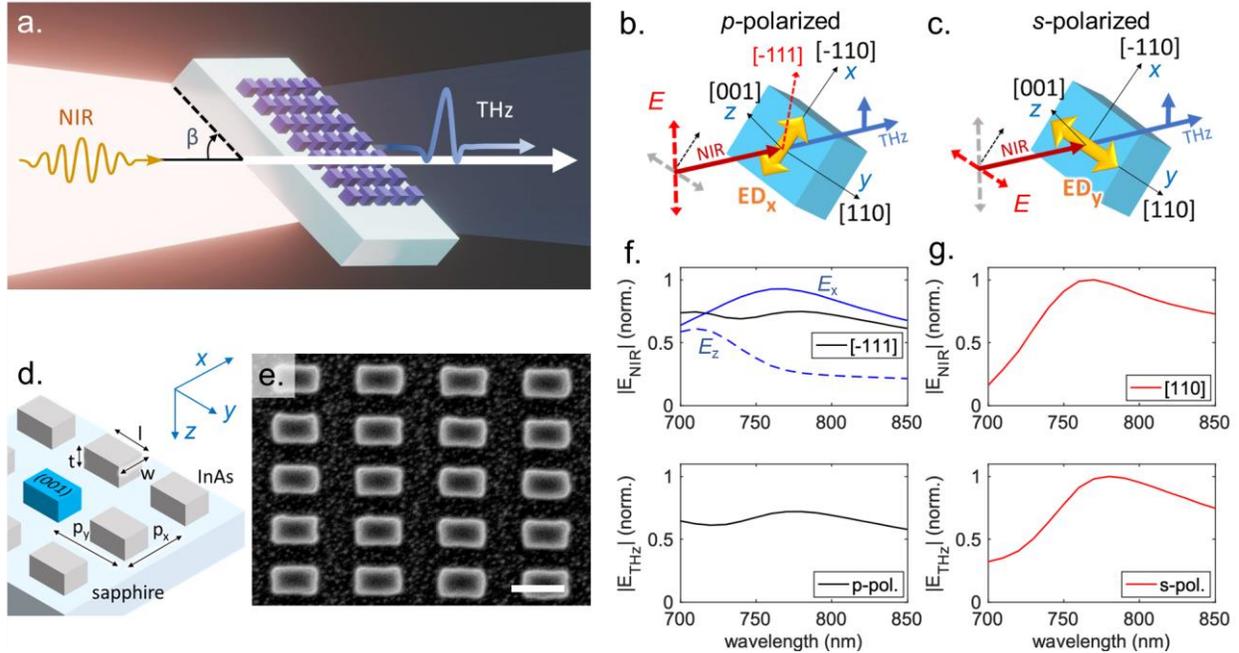

**Figure 1.** InAs metasurface for THz pulse generation. (a) Schematic diagram of THz pulse generation configuration: the InAs metasurface is tilted by angle $\beta = 45$ degrees with respect to the optical axis (shown with the white arrow). (b, c) Schematic diagrams of the InAs resonator showing its orientation and crystallographic axes with respect to the incident excitation electric field vector (red, dashed) for *p*-polarized (b) and *s*-polarized (c) excitation. The excited ED modes are represented using yellow arrows, and the electric field of generated THz pulses is shown in blue. (d) Metasurface design with geometrical parameters of the InAs resonators and their arrangement: w = 160 nm, l = 260 nm, t = 130 nm, $p_x$ = 300 nm, $p_y$ = 480 nm. (e) Scanning electron microscope image of the fabricated metasurface (scale bar: 300 nm). (f, g) Simulated electric field magnitude ($E_x$, $E_z$, $E_{[-111]}$, and $E_{[110]}$) at the resonator center (top row) and estimated *p*-polarized THz pulse amplitude (bottom row) as functions of the optical excitation pulse wavelength. The electric field magnitude spectra and the THz pulse magnitude spectra are normalized to the respective maximum values obtained for *s*-polarized excitation (g).

We numerically simulated the electric field distribution in the metasurface for the 45 degree excitation with 100 fs pulses using the finite difference time domain (FDTD) method. Figure 1f and g show spectra of components $E_x$ and $E_y = E_{[110]}$ of the electric field at the resonator center. The $E_x$ and $E_y$ spectra represent the excitation of the ED modes, $ED_x$ and $ED_y$, and they show clear peaks centered at 770 nm. We note that due to high absorption in InAs, Mie modes in the InAs resonator are broad and in addition to the $ED_x$ and $ED_y$ modes, the 770 nm excitation also couples to the $ED_z$ mode (see Figure 1f) and $MD$ modes. Nevertheless, the $ED_x$ and $ED_y$ modes are the two dominant modes in the resonators excited with $\lambda = 770$ nm (see SI for detailed analysis).

The FDTD simulations within the entire resonator volume allows us to account for the fields of all excited modes, and we can calculate the overall induced non-linear polarization directly from the electric field distribution using the $\chi^{(2)}$ tensor. A projection of that polarization on the vertical axis is used to represent the generation efficiency in the forward direction for *p*-polarized THz pulses (see SI for details). In Figure 1, we show the THz generation efficiency for the *s*- and *p*-polarized 100 fs excitation and correlate it with the spectra of excited modes in the metasurface: the generation efficiency for *s*-polarized excitation follows closely the electric field spectrum for the $E_y$ component ($E_y = E_{[110]}$, Figure 1g), whereas for *p*-polarized excitation, the generation efficiency follows approximately a combined spectrum of the $E_x$ and $E_z$ components, or the spectrum of electric field projection on the [-111] axis, $E_{[-111]}$ (Figure 1e). The similarity between the calculated THz generation efficiency spectra and the electric field components indicate that the THz generation process indeed can be attributed to the excitation of resonator modes, in particular the $ED_x$ and $ED_y$, and the $ED_z$ mode to a lesser degree.

We fabricated the designed metasurface using a 130 nm thick layer of (100) InAs transferred on a sapphire substrate (see Figure 1 and SI for details). We then tested the fabricated metasurfaces for generation of THz pulses using 100 fs optical pulses from the Ti:Sapphire laser (see Figure 2a and SI for details).

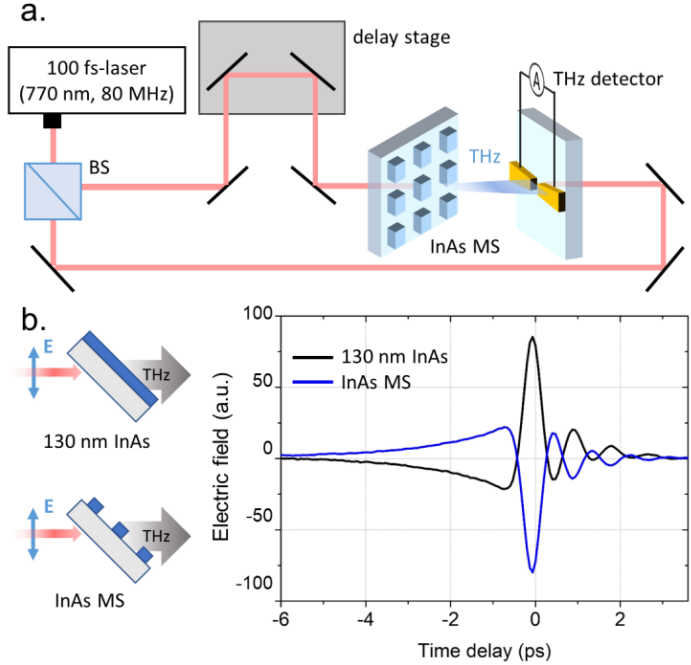

**Figure 2.** THz pulse generation in InAs metasurface. (a) Schematic diagram of the THz emission time-domain spectroscopy setup. (b) Waveforms of THz pulses generated in the InAs metasurface (blue) and in the uniform 130 nm thick InAs layer (black) for *p*-polarized optical excitation.

Figure 2b shows a waveform of the generated THz pulse for the metasurface excited with a *p*-polarized optical beam. For reference, we also show a waveform of the THz pulse emitted from a uniform (unpatterned) InAs layer of the same thickness (130 nm). The experimental results show a striking effect: the metasurface and the uniform InAs layer emit THz pulses of opposite polarities, while their waveforms and amplitudes are practically identical. The opposite polarity corresponds to a π-phase shift for all generated THz frequencies and therefore the metasurface and the uniform InAs layer can serve as two binary-phase elements. We will discuss an experimental demonstration of such a metasurface later; first, however, we briefly comment on THz pulse generation mechanisms.

The polarity flip in Figure 2b indicates that dominant THz generation mechanisms are different in the InAs metasurface and in the uniform InAs layer. To investigate the underlying mechanisms, we characterized the radiated THz pulses for different excitation polarizations. This experiment allows us to differentiate between the photocurrent mechanisms and the optical rectification mechanisms.[35] We find that THz pulses generated from the metasurface strongly depend on the incident polarization angle $\theta$ (Figure 3a). While the pulse shape remains practically unchanged for all excitations polarizations, the polarity changes and the amplitude variation can be described by a function $E_{THz} = ( -A \cos( 2\theta ) + a )$, where $A \gg a$. Almost the same pulse amplitude is observed for the *s*- and *p*-polarized excitations, albeit with opposite THz pulse polarity. Figure 3b summarizes the peak-to-peak THz pulse amplitude as a function of the excitation polarization.

This behavior is indicative of the optical rectification effect. To verify that, we estimated the generated THz pulse amplitude for a varying polarization angle $\theta$ using the second-order nonlinear tensor and numerically simulated electric fields in the metasurface volume (see SI for details). We find that the calculations describe the experimentally observed polarization dependence closely. We emphasize that there is a good agreement between the experiment and the calculations for the metasurface without including any photocurrent effects.

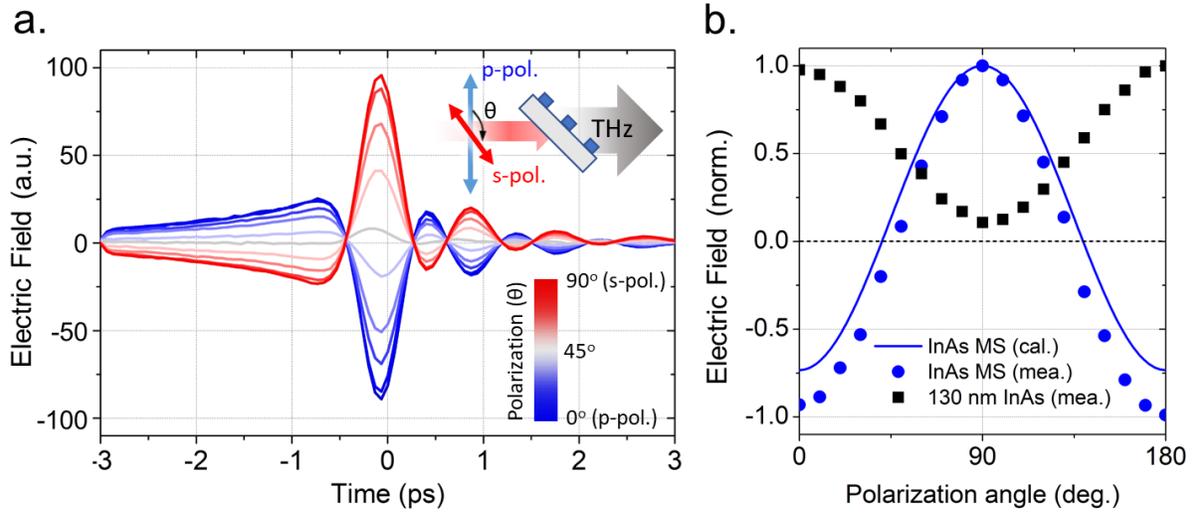

**Figure 3.** THz pulse dependence on excitation polarization. (a) Time-domain waveforms of THz pulses generated in the InAs metasurface for varying polarization of the excitation beam. (b) Normalized peak-to-peak THz pulse amplitude as a function of the excitation polarization angle, $\theta$. Circles show experimentally measured values, and the solid blue line shows the calculated THz pulse generation efficiency due to optical rectification in the InAs resonator volume (excitation wavelength: 770 nm); squares show normalized peak-to-peak amplitude for THz pulses generated in the uniform 130 nm InAs layer.

In contrast, the polarization dependence in the uniform InAs layer is drastically different from the metasurface and we observe no change in the pulse polarity (Figure 3b). Unlike the polarization dependence for the metasurface, the result for the uniform InAs layer cannot be explained by the volume optical rectification effect only. It requires a sizable contribution of a polarization independent (or weakly dependent) mechanism, such as the photo-Dember effect and photocurrents due to surface fields. In fact, the polarization dependence observed for the 130 nm InAs layer is similar to that for bulk InAs crystals.[37] It suggests that THz pulse generation in the uniform InAs layer is a superposition of the photocurrent contributions and the optical rectification contribution, with constructive interference for *p*-polarized excitation and destructive interference for *s*-polarized excitation.

The lack of the photocurrent contribution in the metasurface case is not unexpected because each resonator is small, symmetric and electrically isolated. The photoexcited carriers are excited uniformly within the resonator and they can re-distribute themselves within the small volume sufficiently quickly, resulting in zero net photocurrent and suppression of the photo-Dember effect. Furthermore, the built-in fields at the surfaces are likely to produce opposing photocurrents at each pair of opposing faces of the resonator, also suppressing the net photocurrent. It is unexpected, however, that the polarity of generated THz pulses in the metasurface is opposite to that in the continuous InAs layer, while their absolute amplitudes are similar. In this work, we center on exploiting the polarity effect for THz wavefront engineering and note that further investigation of the underlying mechanisms in InAs metasurfaces could unveil additional schemes for controlling the THz pulse generation process.

We now compare the optical-to-THz conversion efficiency of the InAs metasurface to a standard THz emitter, a 1 mm thick (110) ZnTe crystal.[44] The amplitude of THz pulses generated in the ZnTe crystal is plotted in Figure 4 as a function of peak excitation intensity $I_{opt}$ for reference. The amplitude increases linearly with the intensity following the well-established dependence.[44]

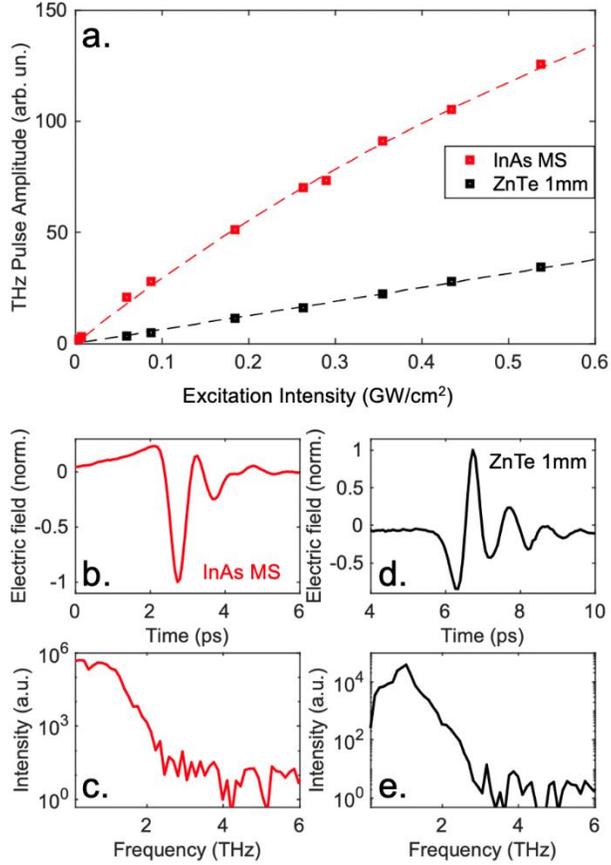

**Figure 4.** THz pulse generation efficiency in InAs metasurface. (a) THz pulse amplitude as a function of peak excitation intensity for the InAs metasurface (red) and 1 mm thick (110) ZnTe crystal (black). Dashed lines show a linear dependence fit to the data for ZnTe (black) and the linear dependence with saturation described in the text. THz pulse waveforms measured for the highest tested excitation intensity (~0.54 GW/cm$^2$) are shown in (b, d) and their Fourier spectra are shown in (c, e).

We then placed the InAs metasurface (at an angle of incidence of 45 degrees, as shown in Figure 1) instead of the ZnTe crystal and recorded THz pulses generated in the metasurface. The pulse amplitude was 3-4 times higher than that for the ZnTe emitter over the entire range of excitation intensities. The experiment demonstrates that the InAs metasurface provides one of the highest optical-to-THz conversion efficiencies.[24,35] Using a literature value of conversion efficiency of ~1.0 × 10$^{-6}$ for a 2 mm thick ZnTe crystal (at 0.133 GW/cm$^2$ of pump excitation), we can estimate that the optical-to-THz conversion efficiency in the InAs metasurface is ~$10^{-5}$ for 0.5 GW/cm$^2$ of excitation (see details in SI).[44]

We note that the functional dependence of THz pulse amplitude on the peak excitation intensity $I_{opt}$ for InAs metasurface is sub-linear. It can be described well with a saturation function: $E_{THz} \sim I_{opt} / (I_{opt} + I_{sat})$, where $I_{sat}$ = 1.5 GW/cm$^2$ is the saturation parameter. This value is comparable with other studies for the optical rectification of InAs.[45] The saturation can be attributed to the increased carrier scattering at high optical fluences and charge screening by the photogenerated carriers in InAs.[3] Nevertheless, typical THz generation systems driven by Ti:Sapphire lasers operate with the excitation intensities below the 1 GW/cm$^2$ level, and therefore the InAs metasurface can provide an efficient solution for THz pulse generation without noticeable saturation. Even for amplified laser sources, which reach higher intensities, we still expect a good conversion efficiency for the InAs metasurface as most known THz sources, including ZnTe, start exhibiting saturation and damage for excitation intensities above 10 GW/cm$^2$.[28]

Finally, we consider the use of developed InAs metasurfaces in practical THz spectroscopy and imaging systems. Focusing of THz pulses is commonly realized using off-axis parabolic mirrors or plastic lenses. These components however tend to be bulky with relatively long focal lengths. Instead, a binary-phase metasurface can be designed to generate and simultaneously focus THz pulses, as it was demonstrated with metallic metasurfaces.[23-26] Figure 5 illustrates the concept of an InAs metasurface consisting of concentric ellipse-shaped binary-phase zones, which generate THz pulses with opposite polarities. When illuminated by an unfocused optical beam, the metasurface generates THz waves, which interfere constructively and form a focal point along the optical axis. This lens design can easily achieve focusing of THz pulses with a focal lengths *f* in the range of several mm desired for practical and compact THz applications.

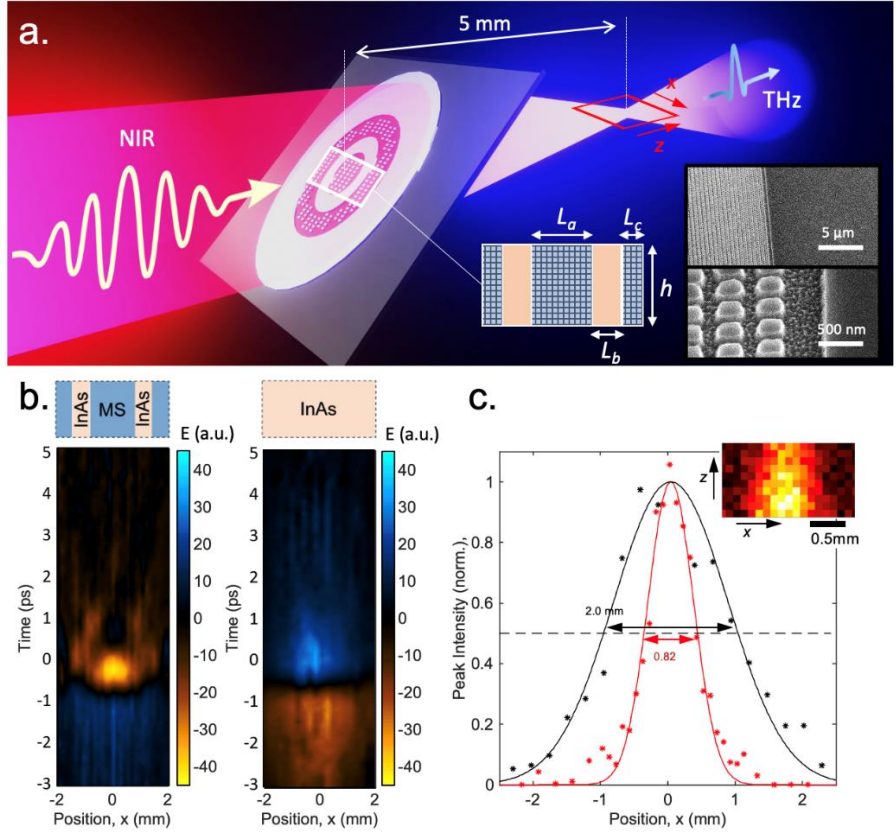

**Figure 5.** Binary phase metasurface for generation and focusing THz pulses. (a) Illustration of the metasurface lens design with insets showing a schematic diagram and SEM images of the sample ($a$ = 1.5 mm, $b$ = 0.7 mm, $c$ = 0.5 mm, and $h$ = 2 mm). (b) Space-time maps of the generated THz field recorded at a distance of $z$ = ~5 mm for the metasurface lens (left) and the uniform InAs layer (right). (c) Normalized THz peak intensity profiles for the metasurface and the uniform InAs layer. Inset: spatial map ($xz$-plane) of THz intensity near the focal point.

We designed a $f$ = 5 mm focusing lens using the InAs metasurface and the uniform 130 nm thick InAs layer as the binary-phase zones (see SI for details). As a proof of concept, we tested a small horizontal section of the designed metasurface lens (2 mm X 4 mm) to verify focusing of the THz pulses in the horizontal plane (along the $x$-axis). A schematic diagram of the lens section and SEM images of the two binary-phase areas are shown in the inset of Figure 5a.

In the experiment, we expanded the 770 nm excitation beam to a diameter of ~ 5 mm to illuminate the entire lens section area. We then mapped the THz field in time and in space (along the *x*-axis) at a distance of ~5 mm from the lens. Figure 5b presents a space-time map of the detected THz field, where one can see a bright spot at *x* = 0 position. It is also possible to recognize three wavefronts of parabolic shape, centered at *x* = – 2, 0, and 2 mm. These wavefronts correspond to THz pulses emitted by the InAs metasurface zones. The bright spot at *x* = 0 represents the region in space where THz emission from all the zones constructively interfere.

In contrast, when the same experiment was repeated with a uniform InAs layer, the map showed no focusing and no parabolic wavefronts. Instead, we observe a THz wave with a flat wavefront and larger diameter (Figure 5b, right panel). The THz amplitude profile simply replicates the intensity profile of the optical excitation beam. To compare the two cases quantitatively, in Figure 5c we show normalized peak THz intensity as a function of position along the *x*-axis. For the metasurface lens, the full width at half maximum (FWHM) is ~0.82 mm (< 3 $\lambda_{THz}$), whereas it is about 2.5 times larger (~2.0 mm) for the uniform InAs layer. To verify focusing along the optical axis, we also mapped the THz pulse peak intensity in the *xz*-plane near the focal point (inset of Figure 5c). The map shows that the sample focuses the generated THz wave along both the transverse (*x*) and longitudinal (*z*) axes, and the maximum intensity is found precisely near the targeted focal point ( *f* = 5 mm). It is important to note that despite the tight field confinement in space-time map in Figure 5b, the focusing varies with the wavelength, as expected for a binary-phase lens. To illustrate the chromatic dispersion, we provide a map of spectral amplitude of the THz field as a function of the wavelength (see SI, Figure S11).

In conclusion, we developed InAs metasurfaces which offer a unique and practical platform for efficient THz pulse generation and simultaneous spatial profile engineering using the binary-phase metasurface concept. Control of the phase in generated THz pulses is possible because different THz generation mechanisms are activated in the nanoscale InAs resonators and the uniform InAs layer. We find that THz pulse generation in InAs resonators is governed by volume optical rectification, whereas in the uniform InAs layer it is likely to be caused by a superposition of surface and volume optical rectification and photoexcited current transients. Polarity and amplitude of THz pulses generated in InAs resonators can be further controlled through the polarization of photoexcitation. It is important to emphasize that the design of the InAs metasurface can be scaled and adjusted for other technologically important wavelengths of excitation (e.g. ~1030 nm and 1550 nm). As a practical example of spatial profile engineering, we demonstrate an InAs metasurface which simultaneously generates and focuses THz pulses with a focal length $f = 5$ mm. This ultimate control of the spatiotemporal structure at the stage of THz pulse generation using nanoscale emitters can eliminate the need for additional, often bulky, inefficient and bandwidth-limiting THz elements, and it opens doors for THz beam engineering, as well as for advanced far-field and near-field THz spectroscopy and imaging applications.

ASSOCIATED CONTENT

**Supporting Information**

The following files are available free of charge.

Theoretical calculation of nonlinearity and corresponding THz generation of InAs, mode decomposition analysis for InAs metasurface, lens design details of structured InAs metasurface, chromatic aberration of the metasurface lens, and methods for fabrication and experiments are included. (PDF)

AUTHOR INFORMATION

**Corresponding Author**

**Oleg Mitrofanov -** University College London, Electronic and Electrical Engineering, London WC1E 7JE, UK; Sandia National Laboratories, Albuquerque, New Mexico 87123, USA; E-mail: o.mitrofanov@ucl.ac.uk

**Hyunseung Jung -** Sandia National Laboratories, Albuquerque, New Mexico 87123, USA; Center for Integrated Nanotechnologies, Sandia National Laboratories, Albuquerque, New Mexico 87123, USA; E-mail: hjung@sandia.gov

**Author Contributions**

The manuscript was written through contributions of all authors.

**Notes**

The authors declare no competing financial interest.

ACKNOWLEDGMENT

This work was supported by the U.S. Department of Energy, Office of Basic Energy Sciences, Division of Materials Sciences and Engineering. LLH was supported by the EPSRC (EP/P021859/1, EP/L015455/1, EP/T517793/1). Metasurface fabrication and characterization were performed at the Center for Integrated Nanotechnologies, an Office of Science User Facility operated for the U.S. Department of Energy (DOE) Office of Science. Sandia National Laboratories is a multi-mission laboratory managed and operated by National Technology and Engineering Solutions of Sandia, LLC., a wholly owned subsidiary of Honeywell International, Inc., for the U.S. Department of Energy's National Nuclear Security Administration under contract

DE-NA-0003525. This article describes objective technical results and analysis. The views expressed in the article do not necessarily represent the views of the U.S. DOE or the United States Government.REFERENCES

(1) Fülöp, J. A.; Tzortzakis, S.; Kampfrath, T. Laser-Driven Strong-Field Terahertz Sources. *Adv. Opt. Mater.* **2020**, 8, 1900681.

(2) Priyadarshi, S. et al. All-Optically Induced Ultrafast Photocurrents: Beyond the Instantaneous Coherent Response. *Phys. Rev. Lett.* **2012**, 109, 216601.

(3) Coté, D. et al. Rectification and Shift Currents in GaAs. *Appl. Phys. Lett.* **2002**, 80, 905.

(4) Planken, P.C.M. et al. Terahertz emission in single quantum wells after coherent optical excitation of light hole and heavy hole excitons. *Phys. Rev. Lett.* **1991**, 69, 3800.

(5) Yue Ma, Y. et al. Recording interfacial currents on the subnanometer length and femtosecond time scale by terahertz emission. *Sci. Adv.* **2019**, 5, eaau0073.

(6) Seifert, T. et al. Efficient Metallic Spintronic Emitters of Ultrabroadband Terahertz Radiation. *Nat. Photon.* **2016**, 10, 483–488.

(7) Maysonnave, J. et al. Terahertz generation by dynamical photon drag effect in graphene excited by femtosecond optical pulses. *Nano Lett.* **2014**, 14, 5797–802.

(8) Jepsen, P. U.; Cooke, D. G.; Koch, M. Terahertz spectroscopy and imaging – Modern techniques and applications. *Laser Photon. Rev.* **2010**, 5, 124-166.

(9) Hsiao, H.-H.; Chu, C. H.; Tsai, D. P. Fundamentals and Applications of Metasurfaces. *Small Methods* **2017**, 1, 1600064.18